\documentclass[preprint2]{aastex}

\usepackage{epsf}
\usepackage{graphics}

\newcommand{\kms}{km s$^{-1}$}
\newcommand{\cts}{counts s$^{-1}$}
\newcommand{\ax}{$\alpha_{\rm X}$}
\newcommand{\aox}{$\alpha_{\rm ox}$}

\newcommand{\cm}{cm$^{-2}$}
\newcommand{\rb}[1]{\raisebox{1.5ex}[-1.5ex]{#1}}
\newcommand{\msun}{$M_{\odot}$}

\newcommand{\pl}{$\pm$}
\newcommand{\nh}{$N_{\rm H}$}
\newcommand{\etal}{{\it et al.}}

\newcommand{\mbh}{$M_{\rm BH}$}
\shorttitle{High redshift quasars}
\shortauthors{Grupe \etal}

\begin{document}


%
%
%


\title{XMM-Newton Observations of High Redshift Quasars
\thanks{Based on observations with XMM-Newton, an ESA Science Mission with
instruments and contribution directly funded by ESA member states and the 
U.S.A.
(NASA).
}
}


\author{D. Grupe\altaffilmark{2}, S. Mathur}
\affil{Astronomy Department, The Ohio State University,
    140 W. 18th Ave., Columbus, OH-43210, U.S.A.}

\email{dgrupe, smita@astronomy.ohio-state.edu}
\altaffiltext{2}{Current address: Astronomy Department, Pennsylvania State
University, 525 Davey Lab, University Park, PA 16802; email:
grupe@astro.psu.edu}

\author{B. Wilkes}
\affil{Harvard-Smithsonian Center for Astrophysics, 60 Garden Street,
    Cambridge, MA 02138; email:belinda@head.cfa.harvard.edu}

\author{and P. Osmer}
\affil{Astronomy Department, The Ohio State University, 140 W. 18th Ave.,
Columbus, OH 43210; posmer@astronomy.ohio-state.edu}




\begin{abstract}
We report on our XMM observations of the high redshift quasars BR
2237--0607 (z=4.558) and BR 0351--1034 (z=4.351). We also report on XMM
observations of 19 other z$>4$ objects available in the public archive
of which 14 were detected.  We find that the optical to X-ray spectral
index \aox~ is correlated with the luminosity density at 2500\AA, but
does not show a correlation with redshift, consistent with earlier
results. Radio loud quasars are brighter and have flatter X-ray slopes
compared to radio-quiet quasars. There is some evidence for the jet
dominated sources to be intrinsically absorbed. The mean intrinsic 2-10
keV power-law slope of the 10 high redshift radio-quiet quasars in our
sample for which a spectral analysis can be performed is
\ax=1.21\pl0.52 (ranging between 0.32 to 1.96),
more like \ax=1.19\pl0.10 found from the ASCA
observations of low redshift Narrow-Line Seyfert 1 galaxies (NLS1s), but 
different from \ax=0.78\pl0.11 found for low redshift Broad-Line Seyfert
galaxies.  The steep X-ray spectral index suggests high Eddington ratios
$L/L_{\rm Edd}$. These observations give credence to the hypothesis of
Mathur (2000) that NLS1s are low luminosity cousins of high redshift
quasars. Comparison with other results from literature indicates that
perhaps most luminous quasars, from low to high redshift, have similarly
steep X-ray spectra suggestive of high Eddington luminosity.
\end{abstract}


\section{Introduction}

High redshift quasars can tell us about the evolution of of central engines
quasars, the star formation history in the early Universe
\citep[e.g.,][]{bromm04, dimatteo04, granato04, die02a}, and the
intergalactic medium between the high redshift quasar and us
\citep[e.g.,][]{sto96, per01}. The high redshift quasars, however, are
typically observed to be faint, making it difficult to analyze them in
detail and to determine their aggregare properties. Quasars are
generally classified as radio-loud (RLQ) and radio-quiet (RQQ) based on
the ratio of their optical to radio luminosity (Kellerman et
al. 1989). Most quasars are radio-quiet, with only about 10\% being
radio-loud. Additionally, a fraction of RLQs, called blazars, have
strong jets pointed towards our line of sight. Many discoveries and
subsequesnt studies of high redshift quasars were of blazars, which are
relatively more bright due to their relativistically enhanced beamed
emission. Naturally, results from these studies cannot be considered
representative of the majority of the radio-quiet quasars. Before the
Sloan Digital Sky Survey \citep[SDSS,][]{yor00} and the {\it Chandra}
Deep Surveys \citep[e.g.,][]{bra01b,giacconi02,barger02,alexander03}
only a handful of high redshift quasars were detected in X-rays.  The
only z$>$4 quasar detected in X-rays before the launch of ROSAT
\citep{tru83} was GB 1508+5714 \citep[z=4.30,][]{mat95} detected by
EINSTEIN.  The first X-ray selected high redshift quasar was RX
J1759.4+6632 \citep[z=4.32,][]{hen94} found in a deep ROSAT Position
Sensitive Proportional Counter \citep[PSPC,][]{pfe86} observation.  Only
one high redshift quasar was discovered during the ROSAT All-Sky Survey
\citep[RASS,][]{vog98}, RX J1028.6--0844 \citep{zic97}.  Other sources
were detected in X-rays, but selected in other wavelength bands,
typically by their radio emission, e.g. GB 1428+4217
\citep[z=4.72,][]{bol00} or at optical wavelengths \citep[e.g.,
Q0000--263, z=4.111,][]{bec94}. Up to the end of 2004 about 85 z$>$4
quasars with X-ray detections are known
\footnote{A complete list of z$>$4
quasars with X-ray detections is given at
www.astro.psu.edu/users/niel/papers/highz-xray-detected.dat,
\citet{bra02b}}.

For the majority of these 85 quasars the number of photons detected is
not sufficient to perform individual spectral analysis. As a result,
derived quantities such as X-ray loudness $\alpha_{\rm
ox}$\footnote{The X-ray loudness is defined by \citet{tan79} as
\aox=--0.384 log($f_{\rm 2keV}/f_{2500\AA}$).}  depend on assumptions
made for spectral shape and absorbing column density. 
Additionally, quasar variability and lack of simultaneous optical and
X-ray observations also contribute to the error on measuring \aox\
(Strateva et al. 2005), leading to conflicting results on the
properties of high redshift quasars. Studies by e.g. \citet{wil94} and
\citet{mat02} showed that \aox~depends on luminosity and not on
redshift (see also Avni \& Tananbaum 1986), while studies by
\citet{bri97} and \citet{bec03} of high redshift quasars claim that
\aox~ increases with redshift. \citet{vig03b} studied large samples
stretching from low-redshift AGN to high-redshift quasars and found
that the trend of \aox~ increasing with redshift is a selection effect
and that \aox~correlates with the luminosity density at 2500\AA~
\citep[see also][]{yuan98}. X-ray spectroscopy is necessary to resolve
these conflicting results.  It has also been claimed by
e.g. \citet{elv94} and \citet{cap97} from ROSAT and ASCA observations
that high-redshift quasars show high intrinsic absorption. Recent
spectroscopic observations by e.g. \citet{fer03} and \citet{gru03}
with XMM contradict these results. Brocksopp et al. (2004) present XMM
spectra of four quasars (two RLQ and two RQQ) at z$=2.96 - 3.77$ and
find that excess absorption is present in one RLQ and one RQQ. The
X-ray properties of the absorbed quasars, however, appear to be
dominated by jet emission.

High-redshift quasars appear to have high mass black holes 
\citep[e.g.,][]{die04, vester04,
 netzer03}. Recent work, e.g. by Merloni (2004), suggests that high
 mass black holes grow rapidly, at high redshift. \citet{mat00}
 suggested that they are similar to low-redshift Narrow-Line Seyfert 1
 galaxies \citep[NLS1s,][]{oster85} which are objects that accrete at
 high Eddington ratios \citep[e.g.][]{bor02, sul00, gru04}. Like
 high-redshift quasars, NLS1s are also thought to be AGN in an early
 evolutionary state \citep{gru96, gru04, mat00}.  It has also been
 found by Mathur el al. (2001) that NLS1s deviate from the well-known
 \mbh-$\sigma$ relation \citep[e.g.,][]{geb00b, ferr00, merr01}
 suggesting that these AGN also have rapidly growing, though smaller
 mass, black holes \citep{gru04c, mat04, mat05}.

To understand high redshift quasars, and to compare them to their low
 redshift cousins, we have initiated a program to obtain X-ray spectra
 of high redshift quasars using XMM-Newton.  The sample consists of
 both radio-loud and radio-quiet quasars to probe differential
 evolution between the two classes, if any.  We have published the
 results of the XMM observations of the radio-loud quasar RX
 J1028.6--0844 and the radio-quiet quasar BR 0351--1034
 \citep{gru04a}. In the present paper we report the observations of
 the radio-quiet quasar BR 2237--0607 and a second observation of BR
 0351--1034.  In addition, we have selected all high redshift
 QSOs$^{2}$ with z$>$4.0 for which public archival data could be
 retrieved from the XMM archive at VILSPA (December 2004).  These 21
 QSOs are listed in Table\,\ref{xmm_qsos}, 16 of which are RQQs,
 representative of the majority of the quasar population and 5 are
 RLQs.

The paper is organized as follows: in \S\,\ref{observe} we describe
the observations and data reduction, followed by a description of the
analysis of the data in \S\,\ref{analysis}. This Section also contains
notes to individual sources.  In \S\,\ref{results} we present the
results of the X-ray observation which will be discussed in
\S\,\ref{discuss}.

Throughout the paper spectral indexes are energy spectral indexes with
$F_{\nu} \propto \nu^{-\alpha}$. The X-ray spectral index \ax~ refers
to the rest-frame energy range 2.0-10.0 keV, except where otherwise noted.
Luminosities are calculated assuming a $\Lambda$CDM cosmology with
$\Omega_{\rm M}$=0.3, $\Omega_{\Lambda}$=0.7, and a Hubble constant of
$H_0$ =75 \kms Mpc$^{-1}$, using the formulae by \citet{hogg99}. In
order to determine the X-ray loudness \aox~ we estimated the flux
density at rest-frame 2500\AA~from rest-frame UV spectra at 1450\AA~
from the literature by assuming UV spectral index $\alpha_{\rm
UV}$=+0.45 as found from composite spectra of  quasars
\citep[e.g.,][]{dan01, die02b}.  The rest-frame flux density at 2 keV
was determined from the XMM pn spectra.  All errors are 1$\sigma$ unless
stated otherwise. We also note that most of the optical and X-ray
observations are separated by years.

\section{\label{observe} Observations and Data Reduction}

\subsection{BR 2237--0607}

The quasar BR 2237--0607 (z=4.558) was observed for the first time in
X-rays by the ROSAT PSPC in May 1993 \citep{kas00}.  The mass of the
black hole of BR 2237--0607 was estimated by \citet{die04} to about
2.9\pl0.8$\times10^{9}$\msun. The black hole mass estimates are made
using the virial theorem and the measurements of CIV emission line
width, optical luminosity and the relationship between the luminosity
and the radius of the broad emission line region \citep{kas00b, vester02}.
\citet{sto96} found several possible intervening damped Ly$\alpha$
absorption systems in the line of sight of BR 2237--0607.  As per the
definition of Kellermann et al. (1989), R=10 marks the radio-loud,
-quiet division, where R=$f_{\rm 5Gz}/f_{\rm 4400\AA}$. While 5
GHz flux of BR 2237$-0607$ is not available in the literature, the
source can be considered radio-quiet based on 1.4 GHz observations
\citep{isaak94}.

BR 2237--0607 was observed by XMM-Newton \citep{jan01} on 2003 May 17
05:52 -- 15:44 (UT) for a total of 32.5 ks with the EPIC pn
\citep{str01} and 34.2 ks with the EPIC MOS \citep{tur01} detectors
using thin filters. A high background flare was present during part of
the exposure, especially towards the end of the observation. We excluded
these times by creating a good time interval (GTI) file accepting only
times when the background count rate of photons with energies $>$ 10 keV
was less than 0.1 \cts.  The GTI screening results in a total observing
time of 23.1 ks for the pn data and of 30.0 ks for the MOS data.  Source
photons in the EPIC pn and MOS were collected in a circle with a radius
of 25$^{''}$ and the background photons in a circle of a radius of
75$^{''}$ close by.  We selected single and double events (PATTERN $\le$
4) for the pn and single, double, triple and quadruple events
(PATTERN$\le$ 12) for the MOS.  Because of the much higher efficiency of
the XMM XRT/pn array compared to the MOS we concentrate on the spectral
analysis of the pn data and use the MOS data for consistency checks
only.

For comparison purposes we also retrieved the ROSAT data
 from the public archive at MPE Garching. The data were 
analyzed by the Extended X-Ray Scientific Analysis System 
\citep[EXSAS,][]{zim98}, version 01APR\footnote{See
http://wave.xray.mpe.mpg.de/exsas/user-guide}.

\subsection{BR 0351--1034}
The quasar BR 0351--1034 (z=4.351) was discovered with the APM
high-redshift quasar survey by \citet{irw91}. It is radio-quiet
\citep{isaak94}.  \citet{sto96} reported that BR0351--1034 was one of
the most unusual sources in their survey of high-redshift APM Quasars
with intervening absorption systems. They found saturated CIV
absorption and a large number of absorption lines associated with
damped Lyman $\alpha$ absorption systems at z=3.633, 4.098, and 4.351.
The source was first detected in X-rays by ROSAT in a 9.1 ks pointed
PSPC observation with 54\pl13 counts \citep{kas00}.  The source has
been observed once before by XMM \citep{gru04a}. Due to high
background radiation throughout, a second XMM-Newton observation was
made in January 2004 and is reported here.

BR 0351--1034 was observed by XMM-Newton on 2004 January 31
 12:02-21:47 (UT) for a total of 32.1 ks with the EPIC pn and 33.8 ks
 with the EPIC MOS detectors using thin filters. Due to high
 background radiation during the last half of the observation, part of
 the pn observation was not usable (the MOS background was much lower,
 so the entire exposure time was useful).  The PN data were screened as
 discussed above, resulting in total of 23.7 ks observing time.
 Source counts of the pn data were collected in a circle with a radius
 of 20$^{''}$ and the background from a near by circular region with a
 radius of 40$^{''}$. Because of its small number of counts
 (Sect. \ref{br0351_res_xmm}) only the EPIC pn data with single and
 double events (PATTERN$\leq$4) were used for the spectral
 analysis. Because of the faintness of the source, proper background
 subtraction is important.  We determined background from a second
 near by region as well, and found the results with the two different
 backgrounds to be consistent within errors.

In order to increase the signal-to-noise ratio of the spectral data,
we merged the GTI screened event files of the two XMM observations of
2002-08-23 \citep{gru04a} and 2004-01-31 (this work) into one event
file using the XMMSAS task {\it merge}.  The data for the spectra were
selected as described above. For the background, a region was chosen
which was observed on the pn CCD4 in both observations. The pn data
were binned with GRPPHA to have at least 15 photons per bin. The
merged data sets results in total observing times of 39.1 ks for the
pn and 53.0 and 53.1 for MOS1 and MOS2, respectively. For consistency
check we also fitted the spectra of the 2002 and 2004 observations
simultaneously using XSPEC.  The analysis and results of the
first XMM-Newton observation have been discussed in
\citet{gru04a}.

\subsection{z$>$4 QSOs with XMM observations}

For all z$>$4 quasars found in the XMM archive we retrieved the
Observational Data Files (ODF) and created the EPIC pn event files
using the XMMSAS task {\it epchain}.  The data were screened for high
background radiation events using the same screening criteria as for
BR0351--1034 and BR 2237--0607.  All z$>$4 QSOs with XMM observations 
are listed in Table\,\ref{xmm_qsos}; multiple observations are
available for 8 of these sources.

All XMM-Newton data were reduced using the XMM-Newton Science Analysis
Software (XMMSAS) version 6.0.0 and the X-ray spectra were analyzed
using XSPEC 11.3.1. The spectra were grouped by GRPPHA 3.0.0 in bins
of at least 20 counts per bin, except otherwise noted. 
The Ancillary Response Matrix and the
Detector Response Matrix were created by the XMMSAS tasks {\it arfgen}
and {\it rmfgen}.  In case of multiple observations, the event files
were merged with the XMMSAS task {\it merge} and the Ancillary
Response Matrices and Detector Response Matrices were combined by the
FTOOLS commands {\it addarf} and {\it addrmf}, respectively.  For the
count rate conversions between different X-ray missions, PIMMS 3.6c was
used.

\section{\label{analysis} Analysis}

\subsection{BR 2237--0607}

\subsubsection{XMM observation}
During the GTI screened observations 381\pl29 source counts were
detected in the pn, 109\pl16 counts in MOS-1, and 143\pl18 in the
MOS-2.  This results in count rates of 0.0165\pl0.0012 counts
s$^{-1}$, (3.67\pl0.56)$\times 10^{-3}$ counts s$^{-1}$, and (4.77\pl0.59)
$\times 10^{-3}$ counts s$^{-1}$ for the pn, MOS-1, and MOS-2, respectively.

\begin{figure*}
\epsscale{1.2}
\plottwo{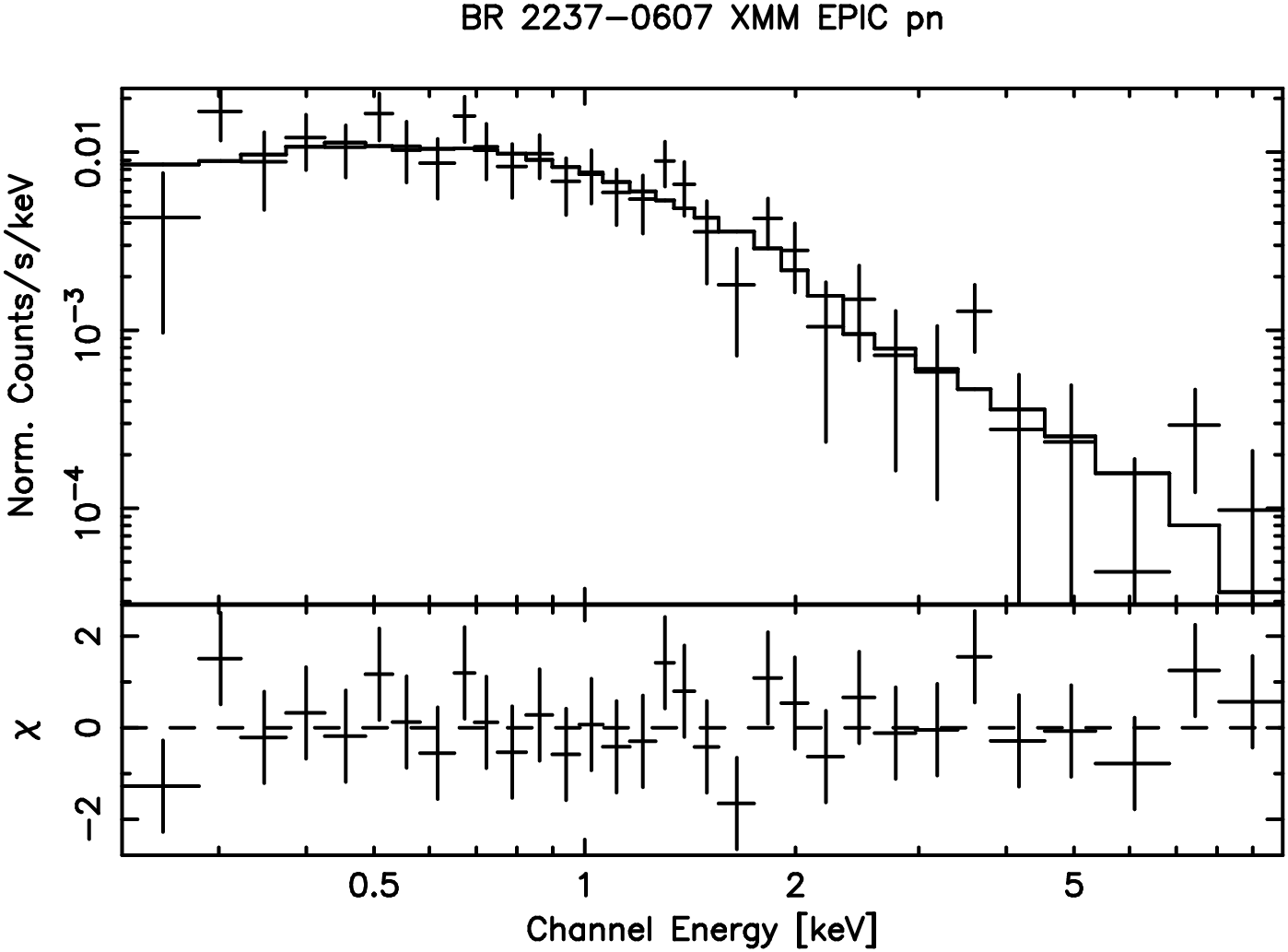}{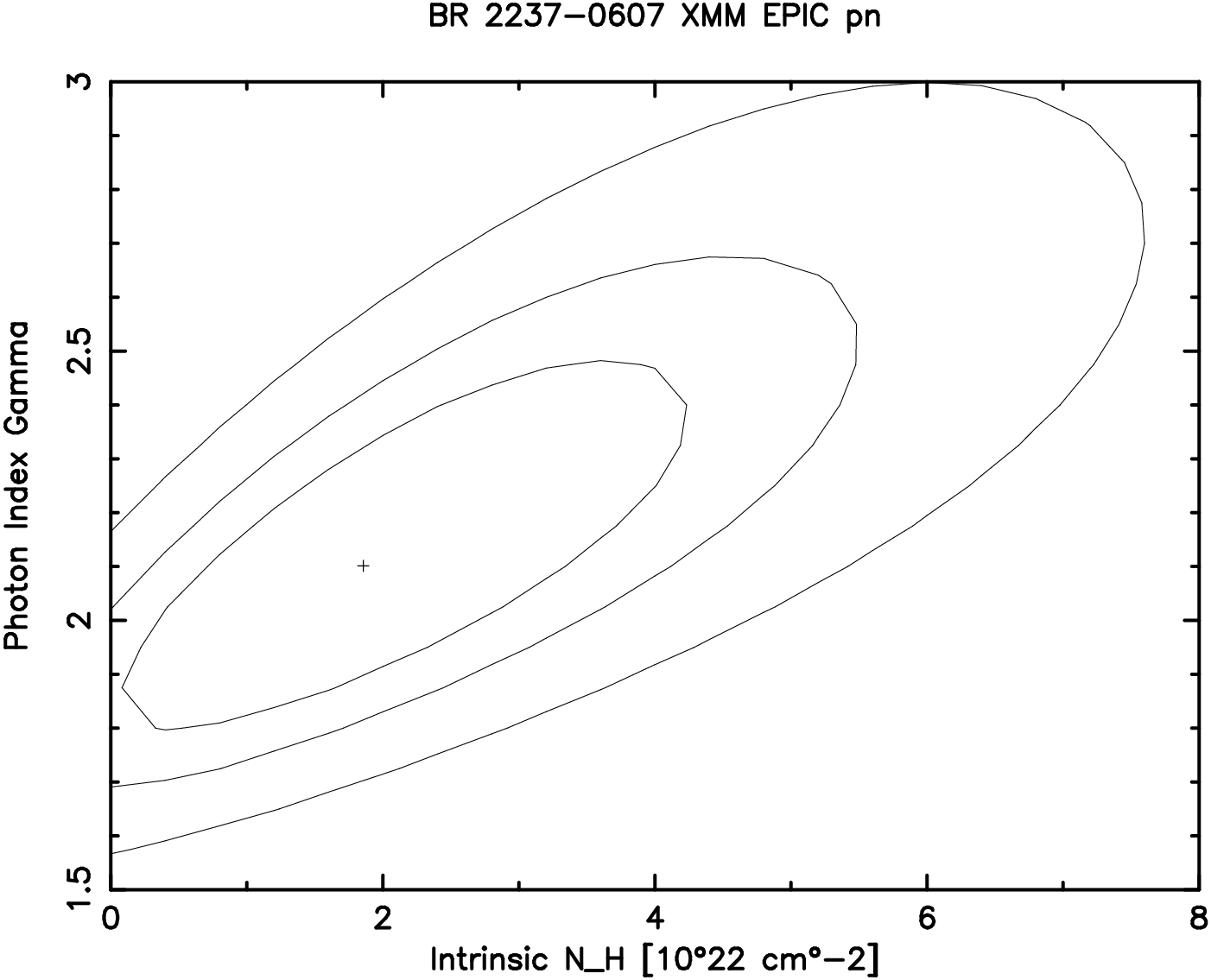}
\caption{\label{br2237_plot} Power-law fit with neutral galactic absorption
(fixed to galactic value) and intrinsic absorption with metal
abundance = solar to the EPIC pn of BR 2237--0607. The left panel shows the fit
to the pn data and the right one the contour plot between the Photon spectral
index $\Gamma$ and the intrinsic column density $N_{\rm H}$.}
\end{figure*}

The results of the spectral analysis are given in
Table\,\ref{br2237_spec}. The data are well-fitted by a single power-law
plus Galactic absorption \nh $=3.84 \times 10^{20}$ cm$^{-2}$
\citep[From the HI maps of ][]{dic90}. While this simple model describes the data well, with no
need for excess absorption, we looked for a signature of excess
absorption in the spectrum to test whether we can rule it out, and
because there exist two intervening absorption systems along the line of
sight of this source. Figure\,\ref{br2237_plot} displays the power-law
fit to the pn data (left panel) and a contour plot between the power-law
slope and intrinsic $N_{\rm H}$ (right panel). Leaving the absorption parameter
free results in a best fit value in excess of Galactic (Figure 1).  The
addition of absorption in excess of Galactic improves the $\chi^2$ by
2.6 for 30 degrees of freedom, corresponding to F-test significance of
F=3.25 and with probability P=0.082. An intrinsic absorber at the
redshift of BR 2237--0607 results in a column density of $N_{\rm
H}=(1.84\pm1.50)\times 10^{22}$ cm$^{-2}$. A Ly$\alpha$ candidate system
at z=4.08 and a Lyman limit system at z=4.28 are observed along the line
of sight to BR 2237$-$0607 \citep{sto96}.  Due to the limited S/N of the
spectrum, however, the redshift of the absorber is not constrained.  If
the absorber redshift corresponds to the 2 intervening absorbers, the
implied column density is $N_{\rm H}=(1.6\pm1.4)\times 10^{22}$
cm$^{-2}$ (z=4.28) and $N_{\rm H}=(1.5\pm1.2)\times 10^{22}$ cm$^{-2}$
(z=4.08). This column density is a factor of about 50 larger than
$N_{\rm HI}=(2.5)\times 10^{20}$ cm$^{-2}$ estimated by \citet{sto96}
for the Ly$\alpha$ system at z=4.08, and would imply unusually large
abundances for the intervening systems. Even though the errors on X-ray
column density measurements are large, the intervening absorption is
excluded at more than $3\sigma$. Therefore, we conclude that the
additional absorption, if present, is intrinsic to the quasar and not to
an intervening system. We reiterate, however, that the possibility of
intrinsic absorption is only a 1$\sigma$ result.
  

The rest-frame unabsorbed 2.0-10.0 keV flux is $2.6\times 10^{-17}$ W
m$^{-2}$ ($2.6\times 10^{-14}$ ergs s$^{-1}$ cm$^{-2}$) which gives a
rest-frame 2-10 keV luminosity of log $L_{\rm 2-10~keV}$=38.68 [W]
(45.68 [ergs s$^{-1}$]).  From the rest-frame UV spectrum 
\citep{die03} and the power law fit to the pn data we derived an X-ray
loudness \aox=1.58\pl0.08.

\subsubsection{Comparison with ROSAT}

From the 9.4 ks ROSAT PSPC observation we measured 37.0\pl14.1 photons
corresponding to a count rate of (3.92\pl1.48) $10^{-3}$ counts s$^{-1}$
in the 0.1--2.0 keV band.  However, this is a factor of about 1.5 lower
than that reported previously by \citet{kas00} who gave a PSPC count
rate = (6.2\pl1.4) 10$^{-3}$ counts s$^{-1}$.  We tried different
background regions in order to search for differences in the background
subtraction, but found this not to be the cause. Our measured count
rate, however, is consistent with the count rate of
(2.94\pl0.67)$10^{-3}$ counts s$^{-1}$ given in the ROSAT source
catalogue and in Vignali et al. (2001), which support the results of our
analysis.  The PSPC count rate is a factor of about 1.6 higher than
expected, based on the pn count rate. However, due to the low number of
PSPC photons and the associated error, the source flux is constant
within 2 $\sigma$.
  
From the ROSAT PSPC data we measured a hardness ratio\footnote{The
hardness ratio is defined as HR=(H-S)/(H+S) with soft band S in the
energy range 0.1-0.4 keV and H between 0.5-2.0 keV} HR=0.37\pl0.34. The
hardness ratio derived from the XMM pn spectrum with a power law model
with Galactic and intrinsic absorption as given in
Table\,\ref{br2237_spec} suggests a much harder spectrum with
HR=0.82. On the other hand, if we use the power law model (\ax=0.87)
with Galactic absorption only, the resulting HR=0.52 which is consistent
with the HR measured from the PSPC. We also checked if additional
information is obtained by subdividing the PSPC energy range to define
two hardness ratios HR1 and HR2\footnote{The HR1 is same as HR defined
above and HR2 is between soft and hard bands of 0.5--0.9 and 0.9--2 keV
respectively.}. We again found that the observed values are more
consistent with a model containing Galactic absorption only rather than
that with excess absorption. The softer spectrum and higher count rate
during ROSAT observation suggest that the BR2237--0607 spectrum is
variable, but the limited S/N of both XMM and ROSAT spectra do not allow
us to draw any firm conclusions.

\subsection{\label{br0351_res_xmm} BR 0351--1034}

The mean count rates during the 2004 January observations are
(5.38\pl0.03) 10$^{-3}$ \cts~ for the EPIC pn, and (2.12\pl0.41)
10$^{-3}$ \cts~ and (1.37\pl0.37) 10$^{-3}$ \cts~ for the EPIC MOS 1
and 2, respectively. This results in total numbers of background
subtracted source counts of 127\pl22 for the pn and 71\pl14 and
44\pl12 for the MOS 1 and 2. The count rates are similar to those
obtained during the first XMM observation on 2002 August 23
\citep{gru04a}.  The merged spectra of the 2002 August and 2004
January observations have a total of 234\pl30, 98\pl18, and 86\pl18
source counts for the pn, MOS1, and MOS2, respectively. The number of
counts in the pn detector are sufficient to perform a basic spectral
analysis; therefore, we focus on the merged pn data only.

Figure\,\ref{br0351_plot} displays powerlaw fits to the 2004 January and the
2002+2004 merged data including contour plots between the intrinsic \nh~and the
photon index $\Gamma$.
Table\,\ref{br0351_spec} summarizes the results of spectral fits to
the EPIC pn data of BR 0351--1034. Since the results of the spectral
analysis of the 2004 January and the merged data from the 2002 August
and 2004 January data are consistent, we discuss the merged data
here. The merged data are consistent with a power law with the
absorption parameter set to the Galactic value
\nh=4.08$\times~10^{20}$\cm, with best fit power-law
slope \ax=0.42$\pm0.17$.  Leaving the absorption parameter free results
in an absorption column at z=0 of 1.0$\pm 0.9\times~10^{21}$\cm. Thus,
even though the best fit column density implies absorption in excess to
the Galactic, this is not a secure result because of the large error on
\nh. If the absorber is placed at the redshift of the quasar the fit
results in an upper limit of the
absorption column of 2.9$\times~10^{22}$\cm. With
the absorbing column fixed at Galactic, the power-law spectral index is
\ax $=0.42\pm0.17$, unusually flat for a radio-quiet quasar. On the
other hand, ROSAT data suggest a steep spectral slope with \ax=3.5 (see
Grupe et al. 2004a for a detailed discussion on comparison with ROSAT
data). Motivated by these results, we fitted the merged data with a
broken power-law model with $\alpha_{X, soft}=3.5$. The resulting fit is
good, with $\alpha_{X, hard}=1.42\pm0.18$, but the data quality is not
good enough to favor one fit over others (see Table\,\ref{br0351_spec}).
For comparison purposes we aslo fitted the 2002 August and 2004 January data
simultaneously in XSPEC. As shown in Table\,\ref{br0351_spec}, we do not find
significant differences between the fits to the merged data set or the
simultaneous fit to the 2 single data sets. 

The absorption corrected rest-frame 2-10 keV flux of the merged
spectrum is log $F_{\rm X}= -17.1$ [W m$^{-2}$ ($-14.1$ [ergs s$^{-1}$
cm$^{-2}$]. This flux converts to an unabsorbed rest-frame 2-10
luminosity 38.13 [W] (45.13 [ergs s$^{-1}$]).  The X-ray loudness is
\aox=1.59\pl0.08 using the rest-frame UV spectrum given in
\citet{sto96}. We note that \citet{kas00} and Vignali et
al. (2001) quote \aox\ of 1.35 and 1.6 respectively.

\subsection{\label{high-z-qsos} High-redshift QSOs with XMM observations}

Here we describe the analysis of all high-redshift quasars with XMM
observations publicly available (December 2004).  All these sources
are listed in Table\,\ref{xmm_qsos}. If enough photons were collected
for a source we performed a power law fit with Galactic absorption
and intrinsic absorption if
needed by the data. The results of these fits and the rest frame 2-10
keV fluxes and luminosities are given in
Table\,\ref{xmm_qsos_spec}. All EPIC pn spectra are shown in
Figure\,\ref{highz_plot}. Several sources have less than 100 detected
counts, resulting in only approximate X-ray spectral slopes; \ax~
derived from these datasets should be treated with caution. The errors
given for \aox~were estimated from the errors in the X-ray spectral
index and the normalization of the spectrum. This is a lower limit of
the error, because it does not take into account errors in the optical
spectrum, e.g. we assume a general optical/UV slope of $\alpha_{\rm
UV}$=0.45 \citep[e.g.,][]{dan01} and \citet{die02b}. Another source of
errors in the optical spectrum is reddening which is difficult to
compute.

\subsubsection{Q 0000--2619}

The XMM observation of the z=4.1 radio-quiet quasar Q 0000--2619
\citep[e.g.][]{bec94} was discussed in detail in \citet{fer03}. The pn
spectrum is well fitted by a single powerlaw with Galactic
absorption. Adding an intrinsic absorber at the redshift of the quasar
  does not improve the fit and the column density can not be
constrained. The best fit power-law slope is \ax$=1.15\pm0.06$. The
X-ray loudness \aox=1.82\pl0.03 was determined from the de-reddened
rest frame UV spectrum given in \citet{bec94} and the rest-frame 2 keV
flux determined from the pn spectrum. This value agrees, within
errors, with the \aox=1.85 given in \citet{fer03} and \citet{bec94},
but differs from \aox=1.65 and \aox=1.71 derived from ROSAT data
reported in \citet{kas00} and Vignali et al. (2001) respectively.
 The ROSAT data presented by \citet{kas00}
contained only 180 counts. While this was a secure detection, the
power-law slope of the spectrum was not as well determined as from the
XMM spectrum, and so \aox\ was not well determined; different
optical slopes also contribute towards different values of
\aox.

\subsubsection{SDSS 0040--0915}

The z=4.98 radio-quiet quasar SDSS 0040--0915 \citep{schneider03} was
serendipitously observed by XMM during a 9ks observation of the
galactic cluster Abell 85 \citep{durret05}.  The 50 photons detected
during this observation only allow a rough estimate of the spectral
slope of the X-ray spectrum. The data were rebinned in groups of at
least 10 counts per bin.  The absorption parameter was fixed to the
Galactic value  which is consistent with the data. The
data can be fitted by a single power law with \ax$=1.43\pm0.34$.  The
X-ray loudness \aox=1.39\pl0.14 was estimated from the rest-frame UV
spectrum given in \citet{schneider03} and the pn spectrum and is in
agreement within the errors with \aox=1.55 given in
\citet{schneider03}.  The steeper X-ray loudness found by
\citet{schneider03} can be explained with their assumption of an X-ray
spectral index \ax=1.0 and the lower count rate and therefore lower flux
in the pn energy band. Also note that \citet{schneider03} use an UV
spectral slope $\alpha_{\rm UV}$=0.79 to extrapolate to the flux density
at rest frame 2500\AA.

\subsubsection{BRI 0103+0032}

This z=4.44 radio-quiet quasar (Smith et al. 1994, Stern et al. 2000)
was observed by XMM for only 4ks.  Only 50 source photons were
detected during the observation, allowing only a rough spectral fit
with a powerlaw model with \ax=1.96\pl0.50 and the absorption
parameter fixed to the Galactic value. The data do not indicate
additional intrinsic absorption. This is the steepest \ax~ known for a
high redshift quasar. Even if the actual value of \ax~ is lower by
1$\sigma$, it is still very steep, comparable to that of NLS1 galaxies
at low redshift (Figure 6). The X-ray loudness \aox=1.41\pl0.36 given
in Table\,\ref{xmm_qsos_spec} was derived from the rest-frame UV
spectrum \citep{die03} and the XMM EPIC pn X-ray data.

\citet{die04} give a black hole mass of 1.0\pl0.3 $10^9$\msun~ and an
Eddington ratio $L/L_{\rm Edd}$=3.5; the black hole mass was determined
using the CIV emission line width and the broad line region
radius--luminosity relation as mentioned above. This high $L/L_{\rm
Edd}$ may explain the steep X-ray spectrum of the source; $L/L_{\rm
Edd}$ and \ax~ have been found to be closely correlated in low-redshift
AGN (Grupe 2004, Williams et al. 2004).  Even though black hole masses
derived from C IV and Mg II widths are highly uncertain, even an order
of magnitude lower $L/L_{\rm Edd}$ is still high, comparable to that of
low redshift NLS1s.

\subsubsection{PSS J0248+1802}
The z=4.422 quasar PSS J0248+1802 was discovered by \citet{ken95} from
a multi-colour survey using the Second Palomar Observatory Sky
Survey. This radio-quiet quasar \citep{stern00} has been a target of
Chandra \citep{vig01} as well as XMM. It was observed by XMM in
February 2003 for a GTI screened time of 7.4 ks
(Table\,\ref{xmm_qsos}) yielding 160 photons. The data were grouped in
bins of at least 15 counts per bin.  A fit to the pn data is
consistent with a single powerlaw with \ax=1.10\pl0.15 with Galactic
absorption.  \citet{vig01} suggested from the Chandra observation that
the source seem to be intrinsically absorbed with a column density
larger than $5\times~10^{23}$\cm.  However, this claim was based on
only 19 source photons and is not supported by XMM data; the best fit
value of intrinsic absorption is an upper limit 
$N_{\rm H}=3\times 10^{20}$\cm (90\%
error).  We estimate the X-ray loudness \aox=1.47\pl0.07 from the
optical/UV spectrum \citep{die03} and the pn spectrum. The black hole
mass was measured by \citet{die04} to be (6.6\pl1.6) $\times~10^9$
\msun\ with an Eddington ratio $L/L_{\rm Edd}$=0.91.

\subsubsection{PMN J0525--3343}
The XMM observations of the blazar PMN J0525--3343 (z=4.40) have been
presented and discussed in detail in \citet{worsley04a} who reported
the presence of a warm absorber in this source, confirming a claim
based on BeppoSAX observations (Fabian et
al. 2001). Table\,\ref{pmn0525_xmm_obs} summarizes the observations
which we merged into one event file. Note that the GTI screened times
given in Table\,\ref{pmn0525_xmm_obs} deviate from the ones given in
Table 1 in \citet{worsley04a}. The reason is a different screening for
the Good Time Intervals (M. Worsley, 2004, priv. comm.).  We did not
use the observations performed during orbits 603 and 608 due to very
high background radiation throughout those orbits.

A single power-law fit to the data with the absorption column density
fixed to the Galactic value \nh=2.21 $10^{20}$ cm$^{-2}$ 
as given in Table\,\ref{xmm_qsos} does not give an acceptable fit
($\chi^2$/dof = 927/601).  The result suggests intrinsic absorption at
the redshift of the quasar. Adding an absorber at the location of the
quasar significantly improved the fit. The best fit results in an
intrinsic cold absorber with $N_{\rm H}=1.6\times 10^{22}$cm$^{-2}$
and an X-ray spectral slope \ax=0.71\pl0.01, which agrees with the
value given by \citet{worsley04a} (0.67\pl0.03); an additional warm
absorber was not required by the data (see residuals to the fit in
Figure 3).

PMN J0525--3343 shows a slight increase in its EPIC pn count rate from
0.311\pl0.006 counts s$^{-1}$ during the February 2001 observation to
0.457\pl0.008 counts s$^{-1}$ during the August 2003 observation.  The
X-ray loudness \aox=1.04\pl0.01 was estimated from the R magnitude
measured from the APM scans given in \citet{per01} and the pn
spectrum. There is a large uncertainty in determine \aox~for PMN
J0525--3343, because we do not know the optical reddening. The
intrinsic absorption column \nh=1.6$\times$10$^{22}$\cm~suggests that
the source may be reddened in the optical. However, the source is
detected in the rest-frame UV, suggesting that the reddening is not
severe. In order to determine the reddening NIR spectra are needed to
measure the Balmer-decrement.

\subsubsection{RX J1028.6--0844}

The z=4.276 radio-loud quasar RX J1028.6--0844 was the only z$>$4 quasar
discovered during the RASS \citep{zic97}. From a long ASCA observation,
\citet{yua00} suggested that the quasar is intrinsically highly absorbed
(\nh=2 $\times~10^{23}$\cm).  However, a 4.4 ks observation with XMM did
not confirm the presence of a strong intrinsic absorber
\citep{gru04a}. Due to the capability of the EPIC pn to observe down
to low X-ray energies of 0.2 keV, \citet{gru04a} measured the
intrinsic absorption column \nh\ of at most a few times
$10^{22}$\cm. From a ten times longer XMM observation \citet{yuan04}
confimed that indeed, the source is not as highly absorbed as claimed
from the ASCA data; the best fit intrinsic cold absorption is
\nh=$2.1^{+0.4}_{-0.3}\times 10^{22}$\cm. \citet{yuan04} argue for an
additional warm absorber with colum density \nh$>2\times 10^{22}$\cm.

\subsubsection{SDSS J1030+0524}

\citet{petric03} found an upper limit of the radio flux at 1.4 GHz
$S_{\rm 1.4 GHz}<$ 61 $\mu$Jy which makes this source radio-quiet.
With a redshift of z=6.28 SDSS J1030+0524 is currently the quasar with
the highest redshift detected in X-rays.  The quasar was
detected by Chandra \citep{mat02, bra02}. It was also observed by XMM
in May 2003 \citep{farrah04} for 67 ks with the pn
(Table\,\ref{xmm_qsos}) resulting in 340 source counts. While a
spectral fit to the pn data with only Galactic absorption provides a
good fit ($\chi^2/\nu$=39/40),
the fit improves significantly
with the addition of intrinsic absorption
of $6.5\pm 3.7\times~10^{22}$\cm ($chi^2/\nu$=32/39). An F-test results in an
F-test value 8.4 with a probability P=0.006. 
The X-ray spectral slope
\ax=1.65\pl0.34 is one of the steepest in the high-redshift quasar
sample presented here. These results are consistent with those of
Farrah et al. (2004) within 2$\sigma$ errors.

\subsubsection{BRI 1033--0327}

The z=4.49 quasar BRI 1033--0327 has been found by \citet{isaak94} to
be radio-quiet.  The 55.7\pl13.6 counts found in the 26.5 ks pn
observation of BRI 1033--0327 allow only a very rough spectral fit to
the data.  The data were binned with at least 10 counts per bin and
are consistent with a simple power law with \ax=1.34\pl0.47 and
Galactic absorption (Table\,\ref{xmm_qsos_spec}). The X-ray loudness
\aox=1.80\pl0.20 was estimated from the rest-frame UV spectrum shown
in \citet{sto96} and the fit to the pn data.  Given the correlation
between \aox\ and optical luminosity (figure 4; Strateva et al. 2005),
large values of \aox\ are expected in highly optically luminous
quasars.  BRI 1033--0327, however, is not very luminous, with $\log
l_{2500\AA}=24.68$ (luminosity density in W Hz$^{-1}$; Table 4),
 making the source unusually X-ray weak. Note, that the error on \aox~is large. 
Moreover, even though the data
are consistent with Galactic absorption only, strong intrinsic
absorption can not be excluded. This would result in underestimating
the intrinsic X-ray luminosity as well as the power-law slope.

\subsubsection{SDSS J1044--0125}

This z=5.74 quasar was discovered by \citet{fan00} and has been found
to be radio-quiet \citep{petric03}.  It was clearly detected by XMM in
May 2000 with 31.7\pl8.5 counts in the observed 0.5-2.0 keV range in
the pn \citep{bra01a}.  The XMM data have been analyzed and discussed
in detail by \citet{bra01a} and \citet{mat01}. The $\approx$30 photons
in this XMM observation do not allow any direct spectral
analysis. \citet{mat01} performed a hardness ratio analysis and argued
that the intrinsic X-ray spectral slope may be steep with \ax=1.5
absorbed by a partially covering absorber.

\subsubsection{RX J1052.4+5719}

This high-redshift quasar (z=4.45) was discovered from ROSAT
observations \citep{schneider98} of the Lockman Hole
\citep{hasinger98}. We used all XMM observations as listed in
\citet{worsley04c} for which RX J1052.4+5719 was in the field-of-view,
using the medium filter (Table\,\ref{rxj1052_xmm_obs}).  From the
screened event files we created a merged event file with a total
observing time of 535 ks (this is the total observing time; the exposure
time at the source is smaller by about a factor of two).  At the
position of the source we found 368\pl52 source photons. Although the
number of photons would suggest good spectral data, this is not the case
for RX J1052.4+5719 due to the high background in the XMM EPIC-pn
detector.  The observed count rate is low, 6.88$\times 10^{-4}$ counts
s$^{-1}$, so the data are dominated by background in the pn detector.
We fitted a single absorbed powerlaw to the data with the absorption
column density fixed to the Galactic value. The fit results in a
relatively steep energy spectral slope \ax=1.58\pl0.30 and does not
require any additional intrinsic absorption.

We estimated an optical/UV -- X-ray spectral index \aox=1.20\pl0.15 from
spectrum given in \citet{schneider98} and the pn spectrum presented
here. The actual, exposure corrected \aox\ is somewhat smaller, about
\aox=1.08\pl0.15. \citet{kas00} reported that RX J0152.4+5719 was not
detected in the NVSS radio survey \citep{cordon98} and gave an upper
limit for the radio loudness of R$<$190.  This makes it difficult to
determine whether RX J1052.4+5719 is a radio-loud or a radio-quiet
quasar; the X-ray spectral index \ax=1.58 suggests that it is
radio-quiet.

\subsubsection{HDF-N objects}

Three spectroscopically confirmed AGNs were detected in the Hubble
Deep Field North (HDF-N) by Chandra: VLA J1236+6213, CXOHDFN
J123647.9+620941, and CXOHDFN J123719.0+621025 \citep{bra01b,
barger02}. This field has also been the target of several XMM
observations (Table\,\ref{rxj1236_xmm_obs}).  The merged event file of
these observations results in a background screened total observing
time of 205 ks (Table\,\ref{rxj1236_xmm_obs}). However, due to the
significantly higher background in the EPIC pn detector compared to
Chandra's ACIS-S, and the much shorter observing time with XMM, none
of the three source could be detected above the background.

\subsubsection{CXOCY J125304--090737}
The z=4.180 quasar CXOCY J125304--090737 was discovered by
\citet{castander03} in a 49.2 ACIS-S observation of the Hickson group
of galaxies HCG 62. It was also observed serendipitously in the field
of view during an 8 ks XMM observation of HCG 62, but we did not
detect the source.  Converting the ACIS-S count rate into EPIC pn
assuming \ax=0.7 and the Galactic absorption column density 3 $\times
10^{20}$ cm$^{-2}$ as given in \citet{castander03} results in 0.0011
pn counts s$^{-1}$. This low count rate in the pn would be dominated
by the detector background. We would have expected to detect only
about 8 source photons during the entire XMM observation, which
explains the non-detection in the pn.

\subsubsection{SDSS J1401+0244}

The z=4.38 quasar SDSS J1401+0244 was serendipitously observed by XMM
in an observation of the galactic cluster A 1835
\citep{schneider03}. We measured 115\pl20 source counts at the
position of the source allowing us for a rough analysis of the X-ray
spectrum. The pn spectrum can be fitted with a very flat X-ray
spectrum with \ax=0.32\pl0.39 and Galactic absorption. As suggested by
\citet{schneider03} there is no evidence of intrinsic X-ray absorption
in SDSS J1401+0244. The radio-loudness of SDSS J1401+0244 is
uncertain. From the FIRST radio survey \citep{becker95} we get a
$5\sigma$ upper limit of 0.95 mJy. With an i-magnitude i=18.6 mag
\citep{schneider03} we can derive an upper limit of the radio-loudness
to R$<$100. The rather flat X-ray spectrum of this quasar suggests
that it is a radio-loud quasar while the X-ray loudness \aox=1.77
\citep{schneider03} suggests it is radio-quiet. Deeper radio
observations of SDSS J1401+0244 are necessary to clarify the radio
nature of this source.  Because we do not have any evidence that the
source is radio-loud we tentatively consider it to be
radio-quiet. Alternatively, the intrinsic X-ray spectrum may be complex,
but appears simple in the low S/N data, leading to underestimating the
true flux.

\subsubsection{GB 1428+4217}

The radio-loud z=4.72 quasar GB 1428+4217 was discovered by
\citet{hook98} and was a target of several X-ray satellites (Fabian et
al. 1997 (ROSAT); 1998 (ASCA); Boller et al. 2000 (ROSAT)).  GB
1428+4217 was actually the last source ever observed by ROSAT in
December 1998 \citep{bol00}.  It was recently observed also by XMM
\citep{worsley04b}.  Spectral fits to the pn data of GB 1428+4217 are
consistent with a power law with Galactic and intrinsic absorption,
confirming earlier results from ROSAT and ASCA (\citet{bol00} and
\citet{fab98}, respectively).  The best-fit power-law slope is
\ax$=0.86\pm0.03$.  While Fabian et al. (2001) argue for the presence
of an intrinsic warm absorber, a fit with cold absorption describes
the XMM data well, with $N_H=2.13\pm0.23$ 10$^{22}$ atoms cm$^{-1}$.  The X-ray
loudness \aox=0.84\pl0.01 was determined from the flux density at
8550\AA (= 1500\AA (1+z)) given in \citet{hook98} and the pn X-ray
spectrum.

\subsubsection{GB 1508+5714}

The z=4.3 radio-loud quasar GB 1508+5714 was discovered in X-rays from
an Einstein observation by \citet{mat95}.  From Chandra ACIS-S
observations \citet{yuan03} and \citet{siemi03} reported the discovery
of an X-ray jet which was also detected in radio at 1.4 GHz with VLA
observations \citep{cheung04}.  GB 1508+1514 was observed twice by XMM
during orbits 443 and 529 on 2002 May 11 and 2002 Oct 30. However, due
to very high background radiation during the second observation, only
the 2002 May 11 observation is useful. The jet is not resolved in XMM
data, so the XMM spectrum is a combined spectrum of the core and the
jet. The flux in jet is only $\sim2$\% of the core flux; therefore the
combined spectrum is dominated by the core component of the quasar
itself. This is clearly seen by the fact that the spectral fit to the
pn data results in a power law with \ax=0.55\pl0.04 and Galactic
absorption, compared to \ax=0.55\pl0.06 found with Chandra data
\citep{siemi03}. Adding an intrinsic absorber to the model does not
improve the fit.  The X-ray loudness \aox=1.07\pl0.02 was determined
from the UV spectrum given in \citet{sto96} and the pn spectrum given
in Table\,\ref{xmm_qsos_spec}.

\subsubsection{RX J1759.4+6838}

The z=4.32 quasar RX J1759.4+6838 was the only X-ray selected z$>$4
quasar that was discovered by ROSAT in a long PSPC observation of the
North Ecliptic Pole \citep{hen94}. The source was serendipitously
observed during two EPIC pn observations of NGC 6552
(Table\,\ref{xmm_qsos}). Assuming an unabsorbed flux of log $F_{\rm
0.5-2.0~keV}=-17$ [W m$^{-2}$] ($-14$ [ergs s$^{-1}$ cm$^{-2}$]) as
given by \citet{hen94} and a powerlaw with \ax=1 and absorption with a
column density $N_{\rm H} = 4 \times 10^{20}$ cm$^{-2}$ 
we would have detected about 40 source counts during the effective 5.9
ks (measured from the combined exposure map). However, the XMM
observation only results in 11.5\pl6.9 pn counts suggesting that
either the spectral shape, normalization or the absorption changes by
on a time scale of about 2 years (rest-frame), resulting in a factor
of 4 change in the observed flux.  \citet{kas00} estimated the
radio-loudness R=30 which makes this source a moderately radio-loud
quasar as per the definition of \citet{kel89}.

\subsubsection{CXOMP J213945--234655}

This z=4.93 quasar was discovered by \citet{silverman02} through the
Chandra Multiwavelength Project (ChaMP, \citet{wilkes01}). The source
was detected on the Chandra ACIS-I detector with 16.7\pl7.7 counts in
the observed 0.3-2.5 keV range during a 41 ks observation. CXOMP
J213945--234655 was also observed by XMM serendipitously during a 12
ks observation of MS 2137--23. We could not detect any source at the
position of CXOMP213945--234655 due to the low flux of the source and
the higher instrumental background of the pn compared to ACIS-S.
There is no source in the NVSS \citep{cordon98} at the position of
CXOMP J213945-234655. Deeper radio observations at the position of
CXOMP J213945--234655 are necessary to clarify whether it is a
radio-loud or radio-quiet quasar. The X-ray loudness \aox=1.52
\citep{silverman02} suggests a radio-quiet nature.

\section{\label{results} Results}

\subsection{BR 2237--0607}

Our spectral analysis of the pn data of BR 2237--0607 has shown that
it does not show significant intrinsic absorption, if present at
all. This result is in agreement earlier findings from XMM data, by
e.g Ferrero \& Brinkmann  (2003, for radio-quiet quasars) and Grupe et
al. (2004a, for radio-loud and -quiet quasars), that high-redshift
quasars are not intrinsically more absorbed than low redshift AGNs.

\citet{die04} estimated the mass of the central black hole to be
2.9\pl0.8 $\times~10^9$\msun~ which corresponds to an Eddington
Luminosity of log $L_{\rm Edd}$=39.4 [W]. The bolometric luminosity
log $L_{\rm bol}$ = 39.7 [W] implies an Eddington ratio $L/L_{\rm
  Edd}$ of about 2 and requires a mass accretion rate of 10 \msun
yr$^{-1}$.  However, even if $L/L_{\rm Edd}$ is lower by factors of
several, it would be still high, comparable to those found among
low-redshift NLS1s \citep[e.g.][]{gru04}.  Typically NLS1s have
Eddington ratios $L/L_{\rm Edd}$ in the order of 1, while Broad Line
Seyfert 1s have $L/L_{\rm Edd}$ about 1 or 2 orders of magnitude
smaller \citep[See Figure 13 in ][]{gru04}.

We can estimate a FWHM(H$\beta$) of BR 2237--0607 from the bolometric
luminosity using the relation by \citet{kas00b}. This results in
FWHM(H$\beta$)$\approx$4500 \kms. This is clearly much larger than the
2000 \kms boundary used to define NLS1s at low redshift. The
definition of ``narrowness'' of FWHM(H$\beta$), should be a function
of luminosity. Only then can we meaningfully expand the NLS1 class to
narrow-line quasar, to include all highly accreting objects. This
result is similar in spirit with recent results from Corbett et
al. (2003) and Shemmer et al. (2004). Using data from 2dF quasar
redshift survey, Corbett et al. have found that the width of the
H$\beta$ line correlates with luminosity, leading to the correlation
of BH mass with luminosity. Similarly, Shemmer et al. have found that
the Eddington luminosity ratio is proportional to the 0.4 power of
luminosity and inversely proportional to the square of H$\beta$
width. Thus, quasars with high Eddington luminosity can have broader
H$\beta$ width, dependent on luminosity.

\subsection{ BR0351--1034}

The spectral analysis of the second XMM observation from January 2004 of
the radio-quiet quasar BR 0351--1034 are in agreement with our earlier
results from the August 2002 observation \citep{gru04a}. In both cases
the data are consistent with a single powerlaw with Galactic absorption.
The merged data sets also yield similar results.  As discussed in \S
3.2, a broken power-law appears to be an appropriate spectral model
for this source (Table 3). Such a steep soft X-ray spectrum with \ax=3.5 would
suggest that BR 0351--1034 is an AGN with a high Eddington ratio
$L/L_{\rm Edd}$ following the correlation between X-ray spectral index
and $L/L_{\rm Edd}$ found among low-redshift AGNs by \citet{gru04} and
narrow line Seyfert 1s by \citet{wil04}. The hard X-ray power-law slope
is $0.42\pm0.18$ with only Galactic absorption and is $0.67\pm 0.21$
allowing for intrinsic absorption (Table 3). However, intrinsic absorption is
not required.

\subsection{z$>$4 quasars}

Figure\,\ref{ax_aox} displays the relation between the X-ray spectral
index \ax~and the X-ray loudness \aox.  As expected from earlier results
by e.g.  \citet{wilkes87} the radio-loud quasars have much flatter X-ray
spectra and smaller \aox\ than the radio-quiet AGNs, similar to that
observed at low redshift.  We did not find a clear correlation between
\ax~and \aox~among the radio-quiet AGN. There is, however, a mild trend
that sources with steeper X-ray spectra tend to be stronger in X-rays,
i.e. have smaller \aox\ ($r_{\rm s}=-0.40, T_{\rm s}=-1.2$, P=0.26). For
our high redshift quasars the observed XMM energy range of 0.2-10 keV
corresponds to a rest frame range of $\approx$1-55 keV.  It is therefore
important to compare the \ax~ values of our sample to the hard X-ray
power-law slopes of low redshift AGNs, as observed by ASCA. The
dashed-dotted lines mark the mean value of the 2-10 keV X-ray slope
\ax~of low-redshift NLS1s (23 sources) and BLS1s (17 sources) of the
ASCA sample of \citet{lei99}. Most of the high-redshift radio-quiet AGN
of our sample have similar steep X-ray slopes as NLS1s, suggesting
similar high Eddington ratios $L/L_{\rm Edd}$. The X-ray spectral slopes of 
radio-quiet AGN with X-ray spectra in our sample ranges between 0.32 to 1.96 and
gives a mean $<$\ax$>$=1.21\pl0.52 which
compares to $<\alpha_{\rm X}>$=1.19\pl0.10 found by \citet{lei99} for
low-redshift NLS1s. The mean \ax~ of the low redshift broad line Seyfert
1s, on the other hand, is 0.78\pl0.11 \citep{lei99}. These values are
similar to the ones found by \citet{bra97} for low redshift
Seyferts. One should be careful, however, in comparing the spectral
shape of our high redshift quasar sample to low redshift Seyfert
galaxies. The X-ray spectra of Seyferts often show complex shapes
including warm absorbers and Compton reflection components. If the
reflection component is not modeled properly, the underlying power-law
slope may be underestimated. As shown in \citet{bra97}, however, the
error introduced is only of the order of $\Delta \Gamma\approx 0.12$,
much larger than the dispersion around mean that we find in our high
redshift sample.

Figure\,\ref{l2500_aox} displays the relation between the rest-frame
luminosity density at 2500\AA, $l_{\rm 2500\AA}$, and X-ray loudness
\aox. As found from earlier studies, e.g. \citet{yuan98b} and
\citet{vig03b}, there is a trend of AGN with higher luminosity
densities at 2500\AA~ to be more X-ray weak then those with lower
$l_{\rm 2500\AA}$.  A Spearman rank order correlation test of the 15
radio-quiet AGN in our sample shows that \aox\ and $l_{\rm 2500\AA}$ are
weakly correlated with a correlation coefficient $r_{\rm s}$=0.28 and a
Student's T test $T_{\rm s}$=1.1 (P=0.14).  The dashed lines in
Figure\,\ref{l2500_aox} mark the mean values for the luminosity
intervals of low-redshift AGN as given in \citet{yuan98}. Thus we see
that the high-redshift quasars in our sample do not show any significant
deviation from the values of low-redshift radio-quiet AGN.

In figure\,\ref{z_aox} we plot \aox~ as a function of redshift. We do
not find a correlation between the two quantities, again consistent
with earlier studies \citep{yuan98b, mat02, vig03b}. Recent work by
Strateva et al. (2005) has also found that the primary dependence of
\aox\ is on luminosity, rather than on redshift. The dotted line
displays the mean value of z$>$2 radio-quiet quasar sample of
\citet{yuan98} observed with ROSAT.  The mean \aox =1.57\pl0.20 of our
15 radio-quiet sources agrees with \aox =1.69\pl0.03 for z$>$2 objects
in the sample of \citet{yuan98}. The mean and the range of \aox values
also are consistent with values in Strateva et al. (2005; see their
figure 11, bottom).

As shown in Figure\,\ref{l2500_lx} radio-loud quasars are brighter by
about two orders of magnitude in the 2-10 keV band than their
radio-quiet cousins for a given luminosity density at 2500\AA.  This
result suggests that the X-ray emission of radio-loud objects is
dominated by beamed emission from the jet.

\section{\label{discuss} Discussion}

In this paper we present XMM detections of 16 high redshift (z$>4$)
quasar, 5 radio-loud and 11 radio-quiet. Spectral shapes were
determined for 4 radio-loud and 10 radio-quiet quasars
(Table\,\ref{xmm_qsos_spec}).  5 other quasars previously detected by
Chandra could not be detected by XMM.  This sample is not complete;
objects were chosen based on their availability in the XMM public
archive. The sample of our own XMM observations was based on previous
X-ray detections, though we focused on observing radio-quiet objects
which are more representative ($\sim 90$\%) of the general quasar
population. Nonetheless, with 14 spectroscopic observations, we are
finally in a position to look for trends in the X-ray properties of
high redshift quasars.

 
 Our results presented here confirm earlier studies by
 e.g. \citet{yuan98} and \citet{vig03b}, that \aox~ depends on
 luminosity (Figure\,\ref{l2500_aox}) and show that there is no clear
 dependence of \aox~ on redshift (Figure\,\ref{z_aox}). The agreement of
 the mean \aox=1.57 of our radio-quiet sources with the radio-quiet
 z$>$2 sample of \citet{yuan98} suggests no evolution of \aox~by
 redshift. Strateva et al. (2005) reached a similar conclusion
 independently.  We find a mild trend of \aox~ decreasing with \ax, but
 this is not a secure correlation.

High redshift radio-loud quasars are more luminous in X-rays and have
flatter X-ray spectra compared to the radio-quiet quasars. This
difference in the two populations is similar to that observed in low
redshift \citep{wilkes87}. Based upon studies of hardness ratios and
flux estimates of a handful of sources Bechtold et al (1994)
 claimed that high redshift radio-loud quasars are more
absorbed than their radio-quiet cousins, suggesting an intrinsic
difference over and above that already present at low redshift. Elvis et
al. (1994) obtained ROSAT spectra of three radio-loud and 3 radio-quiet
quasars. Amoung the radio-loud objects, only one showed $3\sigma$
absorption.  Statistically
significant ($>2\sigma$) intrinsic absorption is present only in 2
out of 5 radio-loud quasars in our sample. Thus, there is some evidence for the
high redshift radio-loud quasars to be more absorbed than the
radio-quiet quasars, even though the sample size is small; the X-ray
spectra of absorbed quasars, however, appear to be dominated by the jet
emission.


We find clear evidence for the 10 radio-quiet quasars with X-ray spectra
in our sample to have steep spectral slopes $<\alpha_{\rm
X}>$=1.23\pl0.48 comparable to those in NLS1 galaxies at low
redshift. If we exclude SDSS J1401+0244, which may be a radio-loud
quasar, the mean and standard deviation change to $<\alpha_{\rm
X}>$=1.33\pl0.38. If we use a flatter slope for BR 0351--1034 (without
intrinsic absorption), then the mean and standard deviation become
$<\alpha_{\rm X}>$=1.20\pl0.48 for all the 10 radio-quiet sources and
$<\alpha_{\rm X}>$=1.30\pl0.44 excluding SDSS J1401+0244. This is an
exciting new result and gives credence to the hypothesis of
\citet{mat00} that NLS1s are low-redshift cousins of high-redshift
quasars, in that they are highly accreting and contain rapidly growing
black holes. Moreover, the high Eddinton luminosities independently
derived for some of our targets, e.g. BRI 0103$+$0032, PSS J0248$+$1802
and BR 2237$-$0607, support the general result.  The mean spectral slope
that we find is consistent with that found by Vignali et al. (2005) and
Shemmer et al. (2005) within the errors. 
A joint fit performed on Chandra and XMM data of
eight radio-quiet z$>4$ quasars led to an ``average'' $\alpha_{\rm
X}=0.97^{+0.06}_{-0.04}$ by Shemmer et al. (2005). Similarly Vignali et
al. performed a joint fit to 48 radio-quiet quasars and got an average
$\alpha_{\rm X}=0.93^{+0.10}_{-0.09}$. We have shown above that the
comparison of spectral shapes between high redshift quasars and low
redshift Seyferts is not significantly affected by spectral complexity
of lower-luminosy Seyferts. It is very useful to make such a comparison
because we have well determined Eddington luminosities only for the
Seyfert population. While we are making progress in determining spectral
shapes, black hole masses and Eddington luminosities in luminous
quasars, these studies are still not done for well defined samples and
are fraught with difficulties.

Shemmer et al. (2004) studied high redshift quasars in the redshift
range $2<z<3.5$ and came to conclusions very similar to ours. They
determined black hole masses using H$\beta$ widths for their sample
objects and found them to have high Eddington luminosities, similar to
NLS1s, and distinct from BLS1s. They also found that such highly
accreting sources, including NLS1 have high metallicities as suggested
by Mathur (2000) (see also Fields et al. 2005). Thus there appears to be
mounting evidence for similarity of properties between NLS1s and high
redshift quasars. Whether all luminous quasars share these properties
remains to be seen.

 In a recent study Piconcelli et al. (2005) performed a systematic
 analysis of low redshift (z$\leq 1.72$) luminous quasars from the
 Palomar-Green (PG) bright quasar sample. They found a mean power-law
 slope of $<\alpha_{\rm X}>=0.89\pm0.11$ for radio-quiet quasars. This
 slope is also statistically consistent with what we find for z$>4$
 quasars, and similar to those of NLS1s.

The above results imply that perhaps {\it most} luminous quasars radiate
at close to Eddington luminosity, based on their X-ray spectral
shape. This may not be a correct interpretation of the observations if
correlation of $\alpha_{\rm X}$ with L/L$_{\rm Edd}$ seen in Seyferts
(e.g. Williams et al. 2004) does not extend to luminous quasars, with
high mass black holes. However, there is some indication in the
literature that such an interpretation may in fact be true. Corbett et
al. (2003) estimated black hole masses for a large number of quasars
over a range of luminosity and found that black hole mass is
proportional to luminosity. This implies that the Eddinton luminosity of
all quasars may be similar. Given the large error in the correlation
equation in Corbett et al., it is not obvious, however, what that
Eddington ratio is. That they all may have high Eddinton luminosity,
similar to NLS1s, is only speculative at this stage (but see Shemmer et
al. 2004); we wish to draw definitive conclusions on the properties of
our own sample only.

The evidence presented in this paper, and other results quoted above,
support the idea that both high redshift quasars and NLS1s accrete at
high Eddington rate. There is, however, one major difference in the two
populations. The high redshift quasars are also very luminous objects,
containing massive black holes of $\approx 10^9$ \msun~ or so
\citep[e.g.,][]{die04, vester04, netzer03}. Thus it appears that while
high mass black holes accrete at high rates and grow most of their mass
at high redshift, the low mass black holes, as in some NLS1s, accrete at
high rate and grow most of their mass at low redshift. This scenario
appears to be consistent with ``anti-hierarchical'' black hole growth
found from the studies of X-ray background \citep[e.g.,][]{meloni04}.

\acknowledgments

We would like to thank Matthias Dietrich for numerous discussions on the
properties of high-redshift quasars and comments and suggestions on the
manuscript.  This research has made use of the NASA/IPAC Extra-galactic
Database (NED) which is operated by the Jet Propulsion Laboratory,
Caltech, under contract with the National Aeronautics and Space
Administration.

 This work was supported in part by NASA grant NAG5-9937.

\clearpage


\begin{figure*}
\epsscale{1.2}
\plottwo{f2a.ps}{f2b.ps}

\plottwo{f2c.ps}{f2d.ps}
\caption{\label{br0351_plot} Power-law fit with neutral galactic absorption
(fixed to galactic value) and intrinsic absorption with metal
abundance = solar to the EPIC pn of BR 0351--1034. The left panels shows the fit
to the 2004 January
pn data and the right ones the fit to the merged data of the 2002 and 2004
observations. The upper panels show the spectral fits to the data and the lower
panels the contours with the Photon spectral index $\Gamma$ and the
intrinsic absorption parameter $N_{\rm H}$.
}
\end{figure*}

\begin{figure*}
\epsscale{2.0}
\plotone{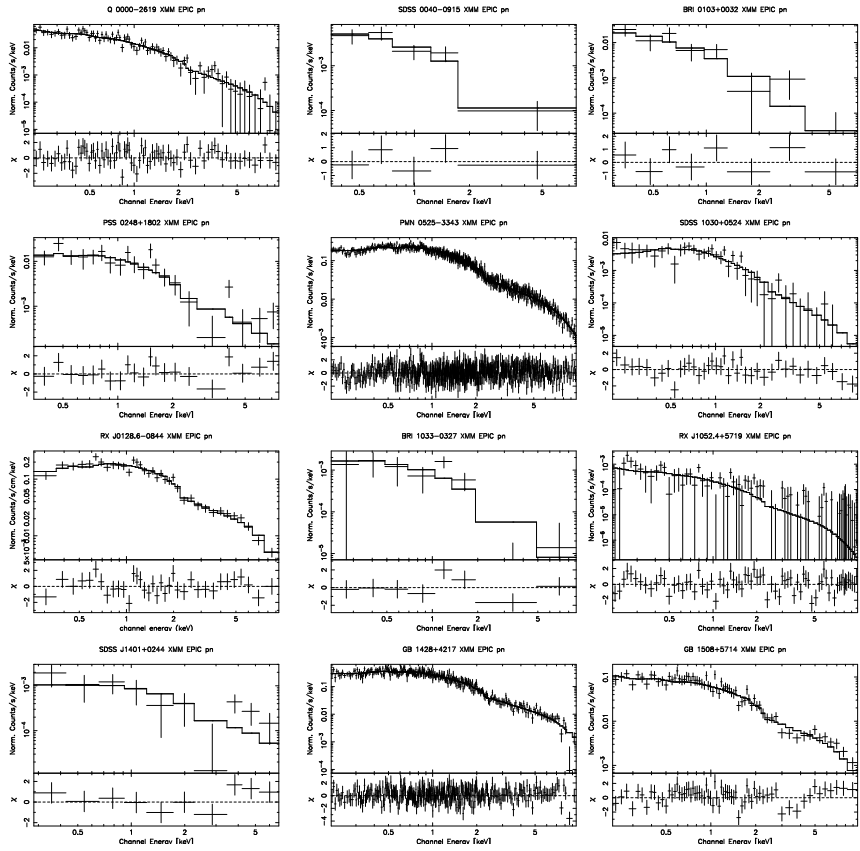}
\caption{\label{highz_plot} Best fits to the XMM EPIC pn spectra of the high
redshift quasars with XMM observations as listed in 
Table \ref{xmm_qsos_spec}.
}
\end{figure*}

\begin{figure}
\epsscale{0.7}
\plotone{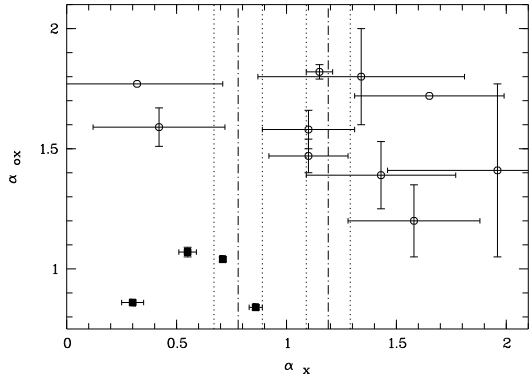}
\caption{\label{ax_aox} X-ray spectral slope \ax~vs. \aox. Symbols are
 as defined in Figure\,\ref{l2500_aox}. The dashed-dotted lines shows
 the mean values of 2-10 keV \ax~of low-redshift BLS1s and NLS1s
 (\ax=0.78 and 1.19 respectively) as given in \citet{lei99}. The dotted
 lines mark the standard deviations around the mean.  }
\end{figure}

\begin{figure}
\epsscale{0.7}
\plotone{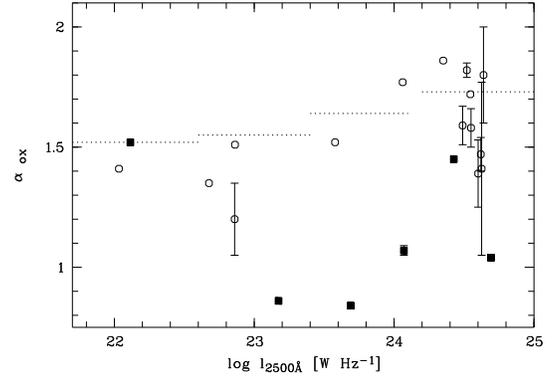}
\caption{\label{l2500_aox} Luminosity density at 2500\AA~vs. \aox. Radio-loud
quasars are displayed as filled squares and radio-quiet quasars as open circles.
For objects without error bars, \aox~was taken from the literature as given in
Table\,\ref{xmm_qsos_spec}.
The dashed lines display the \aox~of low-redshift AGN as given in 
\citet{yuan98}.
}
\end{figure}

\begin{figure}
\epsscale{0.7}
\plotone{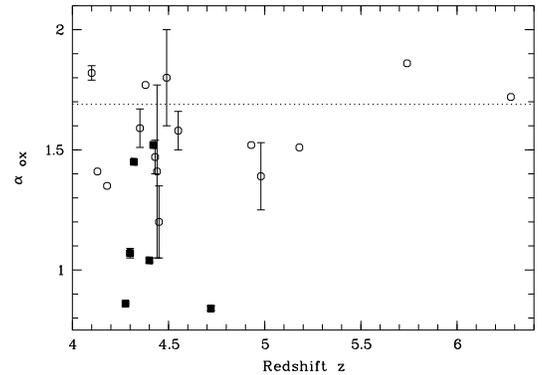}
\caption{\label{z_aox} Redshift z~vs. \aox. Symbols are as defined
in Figure\,\ref{l2500_aox}. The dotted line displays the mean value of Z$>$2
quasars in the sample of \citet{yuan98}.
}
\end{figure}

\begin{figure}
\epsscale{0.7}
\plotone{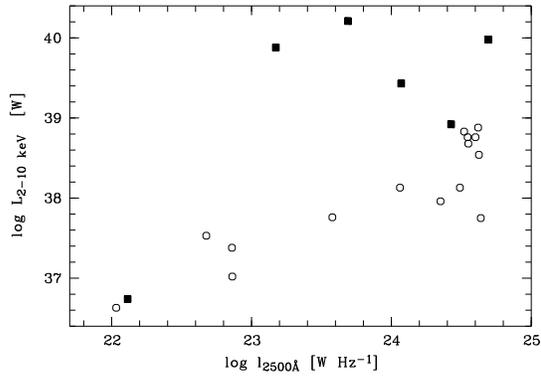}
\caption{\label{l2500_lx} Rest-frame luminosity density at 2500\AA~log 
$l_{\rm 2500\AA}$ vs. rest-frame 2-10 keV X-ray luminosity log $L_{\rm
2-10~keV}$.   
Symbols are as defined in Figure\,\ref{l2500_aox}.
}
\end{figure}

\begin{deluxetable}{rlcccccrrl}
\tabletypesize{\scriptsize}
\tablecaption{z$>$4.0 QSOs with XMM observations
 \label{xmm_qsos}}
\rotate
\tablewidth{20cm}
\tablehead{\colhead{No.} &
\colhead{Object} & \colhead{$\alpha_{\rm J2000}$} & \colhead{$\delta_{\rm
J2000}$} & \colhead{z} & \colhead{XMM Obs-ID} &
\colhead{UT date} & \colhead{$T_{\rm acc}$\tablenotemark{1}} & 
\colhead{Counts}
& \colhead{XMM-Newton reference}  
}
\startdata
1 & Q 0000-2619 & 00 03 22.92 & --26 03 18.7 & 4.100 & 0103060301  & 2002
Jun 25 & 37.8 & 1366\pl46 & \citet{fer03} \\
2 & SDSS 0040--0915 & 00 40 54.65 & --09 15 26.8 & 4.980 & 0065140101  &
2002 Jan 07 & 3.4 & 49\pl8 & \citet{schneider03} \\
3 & BRI 0103+0032 & 01 06 19.20 & +00 48 23.3 & 4.440 & 0150870201 & 2003 Jul
15 & 3.7 & 53\pl11 &  --- \\
4 & PSS J0248+1802 & 02 48 54.30 & +18 02 49.2 & 4.422 & 0150870301 & 2003 Feb 
14
& 7.4 & 161\pl16 & --- \\
5 & BR 0351--1034 & 03 53 46.91 & --10 25 19.0 & 4.351 & \tablenotemark{2}  &
\tablenotemark{2}  & 39.1 & 254\pl30 &  \citet{gru04a} \\
6 & PMN J0525--3343 & 05 25 06.17 & --33 43 05.3 & 4.400 & \tablenotemark{3} & 
\tablenotemark{3} & 72.9 & 28809\pl177  & \citet{worsley04a} \\
7 & RX J1028.6--0844 & 10 28 37.70 & --08 44 23.6 & 4.276 & 0093160701 &
2002 May 15 & 4.4 & 1809\pl44 & \citet{gru04a} \\
8 & SDSS 1030+0524 & 10 30 27.10 & +05 24 55.00 & 6.28 & 0148560501 & 2003
May 22 & 67.5 & 342\pl33 & \citet{farrah04} \\
9 & BRI 1033--0327 & 10 36 23.80 & --03 43 19.3 & 4.49 & 0150870401 &
2002 Dec 20 & 26.5 & 56\pl14 &  --- 
\\
10 & SDSS 1044--0125 & 10 44 33.04 & --01 25 02.2 & 5.740 & 0125300101 &
2000 May  28 &  34.1 & 32\pl9 & \citet{bra01a} \\
11 & RX J1052.4+5719 & 10 52 25.90 & +57 19 07.0 & 4.450 & \tablenotemark{4} &
\tablenotemark{4} & 535.7\tablenotemark{4} 
& 368\pl52 & \citet{worsley04c}  \\
12 & VLA J1236+6213 & 12 36 42.00 & +62 13 31.0 & 4.420 & 
\tablenotemark{5} &
\tablenotemark{5} & 205\tablenotemark{5} & --- &  ---  \\
13 & CXOHDFN J123647+620941 & 12 36 47.90 & +62 09 41.0 & 5.180 & 
\tablenotemark{5}
&  \tablenotemark{5} & 205\tablenotemark{5} & --- &  --- \\
14 & CXOHDFN J123719+621025 & 12 37 19.00 & +62 10 25.0 & 4.130 &
\tablenotemark{5} & \tablenotemark{5} & 205\tablenotemark{5} & --- &  --- \\
15 & CXOCY J125304--090737 & 12 53 04.00 & --09 07 37.0 & 4.180 & 0112270701 
& 2003 Jan 15 & 8.1 & --- &  ---  \\
16 & SDSS J1401+0244 & 14 01 46.53 & +02 44 34.7 & 4.380 & 0098010101 & 2000 
Jun
28 & 32.7 & 115\pl20 & \citet{schneider03} \\
17 & GB 1428+4217 & 14 30 23.78 & +42 04 26.3 & 4.720 & \tablenotemark{6} 
& \tablenotemark{6} & 14.2 & 8476\pl92 & \citet{worsley04b} \\
18 & GB 1508+5714 & 15 10 02.23 & +57 03 04.9 & 4.300 & 0111260201 & 2002 May 
11
& 9.4 & 1437\pl39 & --- \\
19 & RX J1759.4+6638 & 17 59 27.76 & +66 38 53.6 & 4.320 & \tablenotemark{7} &
 \tablenotemark{7} & 6.9 & 12\pl7 &  --- \\
20 & CXOMP J213945--234655 & 21 39 45.00 & --23 46 55.0 & 4.930 & 0008830101 &
2001 Apr 29 & 11.5\tablenotemark{8} & --- &  --- \\
21 & BR 2237--0607 & 22 39 53.57 & --05 52 20.0 & 4.558 & 0149410401 & 
2003 May 17 & 23.1 & 381\pl29 & This paper
\\
\enddata

\tablenotetext{1}{Accepted GTI screened exposure times given in ks.}
\tablenotetext{2}{Merged event file from observation IDs 0093160201 \& 
0203460201
(orbits 495 \& 759; 2002-08-23 \& 2004-01-31)}
\tablenotetext{3}{Merged event file from the observations as listed in
Table\,\ref{pmn0525_xmm_obs}}
\tablenotetext{4}{Merged event file from the observations as listed in
Table\,\ref{rxj1052_xmm_obs}; Note: the exposure time is not derived from the
exposure map}
\tablenotetext{5}{Merged event file from the observations as listed in
Table\,\ref{rxj1236_xmm_obs}; Note: the exposure time is not derived from the
exposure map}
\tablenotetext{6}{Merged event file from observation IDs 0111260101 \& 
0111260701
(orbits 549 \& 569; 2002-12-09 \& 2003-01-17)}
\tablenotetext{7}{Merged event file from observation IDs 0112310401 \& 
0112310801
(orbits 457 \& 523; 2002-06-08 \& 2002-10-18)}
\tablenotetext{8}{Note: the exposure time is not derived from the
exposure map}

\end{deluxetable}

\begin{deluxetable}{ccccr}
\tablecaption{Spectral Fit parameters to the EPIC pn data of BR 2237--0607
 \label{br2237_spec}}
\tablewidth{0pt}
\tablehead{
&  
 \colhead{$N_{\rm H, gal}$} & \colhead{$N_{\rm H, intr}$} \\
 \colhead{\rb{XSPEC Model}} & 
\colhead{10$^{20}$\cm}  & \colhead{10$^{22}$\cm}   
& \colhead{\rb{$\alpha_{\rm X}$}} &
 \colhead{\rb{$\chi^2$ (DOF)}} 
}
\startdata
1 & 8.89\pl4.32 & --- & 1.17\pl0.29 & 21.1 (29) \\
1 & 3.84 (fix) & --- & 0.84\pl0.14 & 26.1 (30) \\
2 & 3.84 (fix) & 1.84\pl1.50 & 1.10\pl0.23 & 23.1 (29) 
\enddata

(1) Power-law with Galactic absorption; 
(2) Power-law with Galactic absorption, and redshifted neutral 
absorption at z=4.558;

\end{deluxetable}

\begin{deluxetable}{lcccccr}
\tablecaption{Spectral Fit parameters to the EPIC pn data of BR 0351--1034
 \label{br0351_spec}}
\tablewidth{0pt}
\tablehead{
&  &
 \colhead{$N_{\rm H, gal}$} & \colhead{$N_{\rm H, intr}$} \\
 \colhead{Obs. Date} & \colhead{\rb{XSPEC Model}} & 
\colhead{10$^{20}$\cm}  & \colhead{10$^{22}$\cm}   
& \colhead{\rb{$\alpha_{\rm X, soft}$}} &
\colhead{\rb{$\alpha_{\rm X, hard}$}} &  \colhead{\rb{$\chi^2$ (DOF)}} 
}
\startdata
2004 Jan & 1 & 15.3\pl14.5 & --- & --- & 0.81\pl0.61 & 6.8 (14) \\
& 2 & 4.08 (fix) & --- & --- & 0.37\pl0.26 & 7.9 (15) \\
& 3 & 4.08 (fix) & 5.34\pl4.30 & --- & 0.75 (fix) & 6.5 (15) \\ \\
Merged 2002+2004 & 1 & 10.26\pl8.51 & --- & --- & 0.69\pl0.37 & 22.5 (25) \\
& 2 & 4.08 (fix) & --- & --- & 0.42\pl0.17 & 23.5 (26) \\
& 3 & 4.08 (fix) & $<$6.3 & --- & 0.67\pl0.30 & 22.0 (25) \\ 
& 4 & 4.08 (fix) & --- & 3.50 (fix)\tablenotemark{1} 
& 0.42\pl0.18 & 22.3 (26) \\ \\
Simultaneous fit\tablenotemark{2} & 1 & 9.65\pl8.09 & --- & --- & 0.82\pl0.39 &
29.8/33 \\
& 2 & 4.08 (fix) & --- & --- & 0.55\pl0.13 & 30.8/34 \\
& 3 & 4.08 (fix) & $<$2.15 & --- & 0.76\pl0.31 & 29.8/33 
\enddata

(1) Power-law with absorption parameter free 
(2) Power-law with Galactic absorption 4.08 10$^{20}$ cm$^{-2}$; 
(3) Power-law with Galactic absorption, and redshifted neutral 
absorption at z=4.351; 
(4) Broken power-law with Galactic absorption

\tablenotetext{1}{$\alpha_{\rm X.soft}$ refers to the soft X-ray spectral slope
that is suggested by fits to the ROSAT data, that suggest a break at 0.45 keV
\citep{gru04a} 
}
\tablenotetext{2}{Simultaneous fit to the 2002 August and 2004 January data in
XSPEC}

\end{deluxetable}

\begin{deluxetable}{rlcccccccccl}
\tabletypesize{\scriptsize}
\tablecaption{Results of power law fits with Galactic (and intrinsic)
absorption to the XMM pn
data of z$>$4.0 QSOs as listed in Table\,\ref{xmm_qsos}. Fluxes and 
luminosities
are given in the rest-frame.
 \label{xmm_qsos_spec}}
\rotate
\tablewidth{23cm}
\tablehead{\colhead{No.} &
\colhead{Object} & \colhead{RL/RQ\tablenotemark{1}}
& \colhead{$N_{\rm H, gal}$\tablenotemark{2}} 
& \colhead{$N_{\rm H, intr}$\tablenotemark{2}} 
& \colhead{$\alpha_{\rm X}$} & \colhead{$\chi^2/\nu$} &
\colhead{log $F_{\rm 2-10}$\tablenotemark{3}} 
& \colhead{log $L_{\rm 2-10}$\tablenotemark{3}} 
& \colhead{log $l_{\rm 2500\AA}$\tablenotemark{3}}
& \colhead{\aox}
& \colhead{Comments}
}
\startdata
1. & Q 0000-2619 & RQ &  1.67 & --- & 1.15\pl0.06 & 75/80 & 
--16.33 & 38.83 & 24.52 & 1.82\pl0.03 & Gal * Powl \\
2 & SDSS 0040--0915 & RQ &  3.37 & --- & 1.43\pl0.34 & 2/3 &
--16.59 & 38.76 & 24.60  & 1.39\pl0.14 & Gal * Powl \\
3 & BRI 0103+0032 & RQ &  3.19 & --- & 1.96\pl0.50 & 5/6 &
--16.70 & 38.54 & 24.63 & 1.41\pl0.36 & Gal * Powl \\
4 & PSS J0248+1802 & RQ & 9.18 & --- & 1.10\pl0.18 & 16/13 &
--16.36 & 38.88 & 24.62 & 1.47\pl0.07 & Gal * Powl \\
5 & BR 0351--1034 & RQ &  4.08 & --- & 0.42\pl0.17\tablenotemark{4} & 23/15 &
--17.10\tablenotemark{4}  & 38.13\tablenotemark{4} & 24.49 
& 1.59\pl0.08\tablenotemark{4} & (Gal +Intr) * Powl \\
6 & PMN J0525--3343 & RL & 2.21 & 1.56\pl0.10 & 0.71\pl0.01 & 601/600 &
--15.24 & 39.98 & 24.68 & 1.04\pl0.01  & (Gal + Intr) * Powl \\
7 & RX J1028.6--0844 & RL &  4.59 & 0.07\pl0.01 & 0.30\pl0.05 & 92/84 &
--15.32 &  39.88
& 23.17 & 0.86\tablenotemark{4} & (Gal + Intr) * Powl \\
8 & SDSS 1030+0524 & RQ &  3.20 & 6.51\pl3.74 & 1.65\pl0.34 & 32/39 &
--16.83 & 38.76  & 24.55 & 1.72\tablenotemark{5} & (Gal + Intr) * Powl \\
9 & BRI 1033--0327 &  RQ & 4.79 & --- & 1.34\pl0.47 & 8/6 &
--17.50 & 37.75 &  24.68 & 1.80\pl0.20 & Gal * Powl \\
10 & SDSS 1044--0125 & RQ &  4.19 & --- & --- & --- & --17.54\tablenotemark{6} &
37.96\tablenotemark{6} & 24.35 & 1.86\tablenotemark{6} & XMM detection \\
11 & RX J1052.4+5719 & RQ &  0.56 & --- & 1.58\pl0.30 & 73/73 &
--17.96 & 37.28 & 22.86 & 1.20\pl0.15 & Gal * Powl \\
12 & VLA J1236+6213 & RL &  1.49 & --- & --- & --- & --18.92\tablenotemark{7}  &
36.74\tablenotemark{7} & 22.11 & 1.52\tablenotemark{7} & No XMM detection\\
13 & CXOHDFN J123647+620941 & RQ &  1.48 & --- & --- & --- & --18.38\tablenotemark{7} &
37.02\tablenotemark{7} & 22.86 & 1.51\tablenotemark{7} & No XMM detection\\
14 & CXOHDFN J123719+621025 & RQ &  1.29 & --- & --- & --- & --18.54\tablenotemark{7} &
36.63\tablenotemark{7} & 22.03 & 1.41\tablenotemark{7} & No XMM detection\\
15 & CXOCY J125304--090737 & RQ &  2.96 & --- & --- & --- & --17.65\tablenotemark{8} 
 &
37.53\tablenotemark{8} & 22.68 & 1.35\tablenotemark{8} & No XMM detection \\
16 & SDSS J1401+0244 & RQ &  2.36 & --- & 0.32\pl0.39 & 9/8 &
--17.09 & 38.13 & 24.06 & 1.77\tablenotemark{9} & Gal * Powl \\
17 & GB 1428+4217 & RL &  1.40 & 2.13\pl0.23 & 0.86\pl0.03 & 299/310 &
--15.09 & 40.21 & 23.69 & 0.84\pl0.01 & (Gal + Intr) * Powl \\
18 & GB 1508+5714 & RL &  1.47 & --- & 0.55\pl0.04 & 75/66 &
--15.77 & 39.43 &  24.07 & 1.07\pl0.02 & Gal * Powl\\
19 & RX J1759.4+6638 & RL &  4.23 & --- & --- & --- & --16.29\tablenotemark{10} &
38.92\tablenotemark{10} & 24.43 & 1.45\tablenotemark{10} & XMM detection\\
20 & CXOMP J213945--234655 & RQ &  3.55 & --- & --- & --- & 
--17.59\tablenotemark{11} &
37.76\tablenotemark{11}  & 23.58 & 1.52\tablenotemark{11} & No XMM detection\\
21 & BR 2237--0607 & RQ & 3.84 & 1.84\pl1.50 & 1.10\pl0.23 & 23/29 &
--16.58 & 38.68  & 24.55 & 1.58\pl0.08 & (Gal + Intr) * Powl 
\\
\enddata

\tablenotetext{1}{Radio-loud (RL) and Radio-quiet (RQ) using the definition of
\citet{kel89}. References on the radio-loudness  are
given in Section\,\ref{high-z-qsos}.}
\tablenotetext{2}{Galactic column density $N_{\rm H, gal}$ in units of 
10$^{20}$ cm$^{-2}$, and intrinsic column density
$N_{\rm H, intr}$ in units of 10$^{22}$ cm$^{-2}$}
\tablenotetext{3}{Rest-frame unabsorbed
flux $F_{\rm 2-10~keV}$ in units of W m$^{-2}$,
luminosity$L_{\rm 2-10~keV}$ in units of W, and 
luminosity density at 2500\AA~given in units of W Hz$^{-1}$.
References for the luminosity density at 2500\AA~are given 
for
each individual object in Section\,\ref{high-z-qsos}.}
\tablenotetext{4}{Estimated from the merged 2002 and 2004 spectra}
\tablenotetext{5}{\aox~ from \citet{mat02}}
\tablenotetext{6}{Based on the data given in \citet{bra01a} converted into flux
using PIMMS and \aox~taken from \citet{bra01a}.}
\tablenotetext{7}{Flux and luminosities values from the Chandra ACIS-S
count rates and X-ray slopes given in \citet{vig02} and 
\aox~taken from \citet{vig02}.}
\tablenotetext{8}{Estimated from the Chandra ACIS-S flux and X-ray slope
 given in \citet{castander03} and \aox~taken from \citet{castander03}.}
\tablenotetext{9}{\aox~taken from \citet{schneider03}.}
\tablenotetext{10}{Estimated from the ROSAT PSPC flux and X-ray slope given in
\citet{kas00} and \aox~taken from \citet{kas00}.}
\tablenotetext{11}{Estimated from the Chandra flux given in 
\citet{silverman02} and \aox~from \citet{silverman02}}

\end{deluxetable}

\begin{deluxetable}{cccr}
\tablecaption{XMM EPIC pn Observations of PMN J0525--3343 \citep{worsley04a}
 \label{pmn0525_xmm_obs}}
\tablewidth{0pt}
\tablehead{
\colhead{Orbit} & \colhead{Obs ID} & \colhead{UT Obs. Date} & \colhead{$T_{\rm
obs}$ [ks]\tablenotemark{1}} 
}
\startdata
216 & 0050150101 & 2001 Feb 11 & 9.8  \\
324 & 0050150301 & 2001 Sep 15 & 20.2 \\
583 & 0149500601 & 2003 Feb 02 & 9.2 \\
588 & 0149500701 & 2003 Feb 24 & 9.3 \\
593 & 0149500801 & 2003 Mar 06 & 9.0 \\
598 & 0149500901 & 2003 Mar 16 & 7.6 \\
671 & 0149501201 & 2003 Aug 08 & 7.6 \\
\enddata
\tablenotetext{1}{screened Good Time Intervals}

\end{deluxetable}

\begin{deluxetable}{cccr}
\tablecaption{XMM Observations of RX J1052.4+5719 with the pn and medium filter
 \label{rxj1052_xmm_obs}}
\tablewidth{0pt}
\tablehead{
\colhead{Orbit} & \colhead{Obs ID} & \colhead{UT Obs. Date} & \colhead{$T_{\rm
obs}$ [ks]\tablenotemark{1}} 
}
\startdata
344 & 0022740101 & 2001 Oct 25 & 14.6  \\
345 & 0022740201 & 2001 Oct 27 & 38.6 \\
349 & 0022740301 & 2001 Nov 04 & 31.9 \\
524 & 0147510901 & 2002 Oct 19 & 57.9 \\
525 & 0147511001 & 2002 Oct 21 & 73.8 \\
526 & 0147511101 & 2002 Oct 23 & 53.0 \\
544 & 0147511601 & 2002 Nov 27 & 97.9 \\
547 & 0147511701 & 2002 Dec 04 & 88.5 \\
548 & 0147511801 & 2002 Dec 06 & 79.5
\enddata
\tablenotetext{1}{screened Good Time Intervals}

\end{deluxetable}

\begin{deluxetable}{cccr}
\tablecaption{XMM Observations of VLA J1236+6213, CXOHDFN J123647+6209, and
CXOHDFN J123719+6210
with the pn and thin filter
 \label{rxj1236_xmm_obs}}
\tablewidth{0pt}
\tablehead{
\colhead{Orbit} & \colhead{Obs ID} & \colhead{UT Obs. Date} & \colhead{$T_{\rm
obs}$ [ks]\tablenotemark{1}} 
}
\startdata
264 & 0111550101 & 2001-05-18 & 37.5 \\
264 & 0111550201 & 2001-05-19 & 35.1 \\
268 & 0111550301 & 2001-05-27 & 23.8 \\
271 & 0111550401 & 2001-06-01 & 81.0 \\
725 & 0162160201 & 2003-11-24 & 10.5 \\
731 & 0162160401 & 2003-12-06 &  8.1 \\
735 & 0162160601 & 2003-12-14 &  9.2
\enddata

\tablenotetext{1}{screened Good Time Intervals}

\end{deluxetable}

\end{document}